\documentclass[showpacs,nofootinbib,prd]{revtex4}

\usepackage{graphicx}
\usepackage{dcolumn}
\usepackage{bm}
\newcommand{\sr}{$\cal S$}
\newcommand{\ps}{$\cal P$}
\newcommand{\ve}{$\cal V$}

\newcommand{\oca}{$\cal O$}
\newcommand{\de}{$\cal D$}
\newcommand{\ds}{\displaystyle}
\newcommand{\lsm}{L$\sigma$M}

\newcommand{\refc}[1]{Ref.~\cite{#1}}
\newcommand{\ha}{\frac{1}{2}}

\newcommand{\bma}[1]{\mbox{\boldmath $#1$}}
\newcommand{\be}{\begin{equation}}
\newcommand{\ee}{\end{equation}}
\newcommand{\bea}{\begin{eqnarray}}
\newcommand{\eea}{\end{eqnarray}}
\newcommand{\eqr}[1]{Eq.~(\ref{#1})}
\newcommand{\eqrs}[2]{Eqs.~(\ref{#1}) and (\ref{#2})}

\begin{document}
\title{Ground-State Scalar \bma{\bar{q}q} Nonet: $SU(3)$ Mass Splittings and \\
Strong, Electromagnetic, and Weak Decay Rates}
\author{Michael D.\ Scadron}
\email{scadron@physics.arizona.edu}
\affiliation{Physics Department, University of Arizona, Tucson, AZ 85721, USA} 
\author{George Rupp}
\altaffiliation[Corresponding author]{}
\email{george@ist.utl.pt}
\author{Frieder Kleefeld}
\email{kleefeld@cfif.ist.utl.pt}
\affiliation{Centro de F\'{\i}sica das Interac\c{c}\~{o}es Fundamentais,
Instituto Superior T\'{e}cnico, P-1049-001 Lisboa, Portugal} 
\author{Eef van Beveren}
\email{eef@teor.fis.uc.pt}
\affiliation{Departamento de F\'{\i}sica, Universidade de Coimbra,
P-3004-516 Coimbra, Portugal} 
\date{\today}
\begin{abstract}
By comparing $SU(3)$-breaking scales of linear mass formulae, it is shown that
the lowest vector and scalar mesons all have a $\bar{q}q$
configuration, while the ground-state octet and decuplet baryons are $qqq$.
Also, the quark-level linear $\sigma$ model is employed to predict similar
$\bar{q}q$ and $qqq$ states. Furthermore, the approximate mass degeneracy of
the scalar $a_0$(985) and $f_0$(980) mesons is demonstrated to be accidental. 
Finally, it is shown that various strong, electromagnetic, and weak mesonic
decay rates are successfully explained within the framework of the quark-level
linear $\sigma$ model. 
\end{abstract}
\maketitle
\section{Introduction}
In the quark model, one usually assumes that pseudoscalar (\ps), and vector
(\ve) mesons are $\bar{q}q$, whereas octet (\oca) and decuplet (\de) baryons
are $qqq$ states. However, it is now argued \cite{PDG02} that the light scalar
(\sr) mesons are non-$\bar{q}q$ candidates, in view of their low masses and
broad widths. In this paper, we shall show that the ground-state meson nonets
\ps, \sr, and \ve\ are \em all \em \/$\bar{q}q$, hence including the light
scalars, while the lowest \oca\ and \de\ baryons are $qqq$ states. 

In Sec.~2, $SU(3)$ mass splittings for loosely bound \ve\ and \sr\ states
are shown to have symmetry-breaking scales of 13\% and 18\%,
respectively, using linear mass formulae. We apply the latter formulae to
$qqq$ \oca\ and \de\ states in Sec.~3, leading to $SU(3)$-breaking scales of
13\% and 12\%, respectively. Then in Sec.~4, we employ the quark-level linear
$\sigma$ model (\lsm) to predict similar $\bar{q}q$ and $qqq$ states as in
Secs.~2 and 3. Next in Sec.~5, we study the \sr\ $\bar{q}q$ states and argue
why the \ve\ states have slightly higher masses, on the basis of the
nonrelativistic quark model. Moreover, the approximate mass degenaracy of the
\sr\ $a_0$(985) and $f_0$(980) mesons is shown to be just accidental. 
Finally, in Secs.\ 6, 7, and 8 we successfully determine, in an \lsm\
framework, mesonic decay rates for strong, electromagnetic, and weak 
processes, respectively. In Sec.~9 we summarize our results and draw some
conclusions.

\section{Mass splittings for \bma{U(3)\times U(3)} \ve\ and \sr\ \bma{\bar{q}q}
mesons}

Although meson masses are expected to appear \em quadratically \em \/in model
Lagrangians, while they must appear so for \ps\ states \cite{PS80}, for \ve\
and \sr\ states a Taylor-series linear form for $SU(3)$ mass splittings is
also possible. Thus, consider a Hamiltonian density $H=H(\lambda_0)+
H_{ss}(\lambda_8)$ using Gell-Mann matrices. Then the vector-meson-nonet masses
$m_{\cal V}= \sqrt{2/3} \;\, m_{\cal V}^0 - \, d_{\bar{i}8i} \; \delta m_{\cal V}$ are\footnote{Recall that $d_{0\,i\,j} = \sqrt{2/3} \; \delta_{ij}\;$, $d_{\bar{n}\,8\,n}=1/\sqrt{3}$, $d_{\bar{s}\,8\,n}=d_{K\,8\,K}=-1/(2\,\sqrt{3})$, and $d_{\bar{s}\,8\,s}=-2/\sqrt{3}$.}
\be
\begin{array}{ccccr} 
m_{\rho,\omega}&=&\displaystyle\sqrt{\frac{2}{3}} \;\, m_{\cal V}^0\,-\;\,\displaystyle\frac{1}{\sqrt{3}} \,\;\delta m_{\cal V} &\approx&776\;\mbox{MeV} \; ,\\[4mm]
m_{K^*}&=&\displaystyle\sqrt{\frac{2}{3}} \;\, m_{\cal V}^0\,+\,\displaystyle\frac{1}{2\sqrt{3}}\,\;\delta m_{\cal V}&
\approx&894\;\mbox{MeV} \; , \\[4mm]
m_{\phi}&=&\displaystyle\sqrt{\frac{2}{3}} \;\, m_{\cal V}^0\,+\;\,\displaystyle\frac{2}{\sqrt{3}}\,\;\delta m_{\cal V}&
\approx&1020\;\mbox{MeV} \; ,
\end{array}
\label{vector}
\ee
with $\phi\approx\bar{s}s$. Measured vector masses \cite{PDG02} suggest average
mass splittings
\be
m_{\cal V}^0 \; \approx \; 1048\;\mbox{MeV} \;\;\; , \;\;\; \delta m_{\cal V} \;
\approx \; 141\;\mbox{MeV} \;\;\; ,
\label{mv}
\ee
giving an $SU(3)$-breaking scale of
$\delta m_{\cal V}/m_{\cal V}^0\approx13\%$.

Such considerations can be repeated for axial-vector mesons as well, even
though it is now hard to draw any decisive conclusions, also in view of the
experimental situation. This is why regarding these mesons we limit ourselves
to the following observations. In the case of axial-vector $a_1$ states, we
assume the $f_1$(1420) is mostly $\bar{s}s$, because the PDG
\cite{PDG02} reports $f_1(1420)\to KK\pi, \,K^*K$ as dominant, while $f_1(1285)
\to KK\pi, \,K^*K$ are almost absent. Thus, $f_1(1285)$ is mostly $\bar{n}n$,
like the nonstrange $a_1(1260)$ (with $a_1\to\sigma\pi$ seen, but $a_1\to
f_0(980)\pi$ not seen, because $f_0(980)$ is mostly $\bar{s}s$). 

Also the scalar masses (not incompatible with \refc{PDG02}) predicted from
the \lsm\ discussed in Sec.~4  obey the mass-splitting pattern (for the chiral
limit (CL) in $SU(2)$ and SU(3), see Refs.~\cite{DS95} and
\cite{Delbourgo:1998kg}, respectively)
\be
\begin{array}{ccccccl} 
m_{\sigma_n}&=&\displaystyle\sqrt{\frac{2}{3}} \;\, m_{\!\cal S}^0\,-\;\,\displaystyle\frac{1}{\sqrt{3}} \, \;\delta m_{\!\cal S} & \stackrel{CL}{\rightarrow} & 2\; \hat{m}_{CL} & = & 650\;\mbox{MeV} \; , \\[4mm]
m_{\kappa}&=&\displaystyle\sqrt{\frac{2}{3}} \;\, m_{\!\cal S}^0\,+\,\displaystyle\frac{1}{2\sqrt{3}}\, \;\delta m_{\!\cal S}&\stackrel{CL}{\rightarrow} & 2\,\; \sqrt{\,\hat{m}_{CL} \, m_{s,\,CL}} & = & 780\;\mbox{MeV}  \; , \\[4mm]
m_{\sigma_s}&=& \displaystyle\sqrt{\frac{2}{3}} \;\, m_{\!\cal S}^0\,+\;\,\displaystyle\frac{2}{\sqrt{3}} \, \;\delta m_{\!\cal S} & \stackrel{CL}{\rightarrow} & 2\; m_{s,\,CL} & = & 940\;\mbox{MeV} \; .
\end{array}
\label{scalar}
\ee
Here, $m_{\sigma_n(650)}$ is near the PDG average \cite{PDG02}
$m_{f_0(600)}$, \,$m_{\kappa(780)}$ is near the E791 value \cite{A02}
$797\pm19$ MeV, and $m_{\sigma_s(940)}$ is near the PDG value $m_{f_0(980)}$, which is thus mostly $\bar{s}s$.
The masses from Eqs.~(\ref{scalar}) then give the CL average mass splittings 
\begin{eqnarray}
m_{\!\cal S}^0 & \stackrel{CL}{\rightarrow} & 922 \;\mbox{MeV}\;, \nonumber \\[2mm]
\delta m_{\!\cal S} & \stackrel{CL}{\rightarrow} & 167 \;\mbox{MeV} \;, \label{ms} \\[2mm]
\frac{\delta m_{\!\cal S}}{m^0_{\!\cal S}} & \stackrel{CL}{\rightarrow} & 18\;\%\;\;.\nonumber
\end{eqnarray}

The fact that the $\bar{q}q$ scalars have an $SU(3)$-breaking CL scale of 18\%,
larger than the 13\% scale of \ve\ ground states, further suggests that,
whereas the \ve\ are $\bar{q}q$ loosely bound states, the $\bar{q}q$ \sr\
states (with quarks touching in the NJL scheme \cite{NJL61}) are ``barely''
elementary-particle partners of the tightly bound \ps\ states (discussed in
Sec.~4).

\section{Loosely bound \bma{qqq} baryons}

In this same Taylor-series spirit, the octet (\oca) baryon $SU(3)$ mass
splitting 
$m_{\!\cal O}=m^0_{\!\cal O}-\delta m_{\!\cal O}(d_{ss}d^{\bar{i}8i}+
f_{ss}if^{\bar{i}8i})$, with $d_{ss}+f_{ss}=1$,
predicts (the index $ss$ means semistrong)  
\be
\begin{array}{cclcr} 
m_{N}&=&m_{\!\cal O}^0\,-\,\displaystyle\frac{\delta m_{\!\cal O}}{2\sqrt{3}}\,
(-d_{ss}+3f_{ss})&\approx&939\;\mbox{MeV} \; , \\[4mm]
m_{\Lambda}&=&m_{\!\cal O}^0\,+\,\displaystyle\frac{\delta m_{\!\cal O}}
{\sqrt{3}}\,d_{ss}&\approx&1116\;\mbox{MeV} \; , \\[4mm]
m_{\Sigma}&=&m_{\!\cal O}^0\,-\,\displaystyle\frac{\delta m_{\!\cal O}}
{\sqrt{3}}\,d_{ss}&\approx&1193\;\mbox{MeV} \; , \\[4mm]
m_{\Xi}&=&m_{\!\cal O}^0\,+\,\displaystyle\frac{\delta m_{\!\cal O}}{2\sqrt{3}}
\,(d_{ss}+3f_{ss})&\approx&1318\;\mbox{MeV} \; . 
\end{array}
\label{octet}
\ee
The $(d/f)_{ss}$ ratio can be found from Eqs.~(\ref{octet}) as
\be
\left(\frac{d}{f}\right)_{\!\!\!ss}\;=\;-\frac{3}{2}\:\frac{m_\Sigma-m_\Lambda}
{m_\Xi-m_N} \; \approx \; -0.305 \;\;\;,\;\;\; d_{ss} \; \approx \; -0.44
\;\;\;,\;\;\; f_{ss} \; \approx \; 1.44 \;\;\; .
\label{dfo}
\ee
Thus, Eqs.~(\ref{octet}) predict the average mass splittings
\be
m_{\!\cal O}^0 \; \approx \; 1151\;\mbox{MeV} \;\;\;,\;\;\; \delta m_{\!\cal O}
\; \approx \; 150\;\mbox{MeV} \;\;\;,\;\;\;
\frac{\delta m_{\!\cal O}}{m^0_{\!\cal O}} \; \approx \; 13\% \;\;\;.
\label{mo}
\ee

The $SU(3)$ \de\ baryon masses
$m_{\!\cal D}=m^0_{\!\cal D}+\delta m_{\!\cal D}$ have $m^0_{\cal D}$ weighted
by wave functions
\be
\overline{\Psi}^{(abc)}\Psi_{(abc)} \; = \; \overline{\Delta}\Delta \, + \,
\overline{\Sigma}^*\Sigma^* \, + \, \overline{\Xi}^*\Xi^* \, + \,
\overline{\Omega}\Omega \; ,
\label{dwf}
\ee
and $\delta m_{\!\cal D}$ is weighted by
\be
3\,\overline{\Psi}^{(ab3)}\Psi_{(ab3)} \; = \; \overline{\Sigma}^*\Sigma^* \, +
\, 2\,\overline{\Xi}^*\Xi^* \, + \, 3\,\overline{\Omega}\Omega \; .
\label{dmd}
\ee
Then the $SU(3)$ \de\ masses are predicted (in MeV) to be
\be
\begin{array}{cclcl}
m_\Delta & = & m^0_{\!\cal D} & \approx & 1232 \; , \\[2mm]
m_{\Sigma^*} & = & m^0_{\!\cal D}\,+\,\delta m_{\!\cal D} & \approx & 1385 \; ,
\;\;\; \mbox{with} \;\; \delta m_{\!\cal D} \; \approx \; 153 \; , \\[2mm]
m_{\Xi^*} & = & m^0_{\!\cal D}\,+\,2\delta m_{\!\cal D} & \approx & 1533 \; ,
\;\;\; \mbox{with} \;\; \delta m_{\!\cal D} \; \approx \; 151 \; , \\[2mm]
m_{\Omega} & = & m^0_{\!\cal D}\,+\,3\delta m_{\!\cal D} & \approx & 1672 \; ,
\;\;\; \mbox{with} \;\; \delta m_{\!\cal D} \; \approx \; 147 \; .
\end{array}
\label{decuplet}
\ee
This corresponds to average mass splittings
\be
m_{\!\cal D}^0 \; \approx \; 1232\;\mbox{MeV} \;\;\;,\;\;\; \delta m_{\!\cal D}
\; \approx \; 150\;\mbox{MeV} \;\;\;,\;\;\;
\frac{\delta m_{\!\cal D}}{m^0_{\!\cal D}} \; \approx \; 12\% \;\;\;.
\label{md}
\ee

It is interesting that both loosely bound $qqq$ \oca\ and \de\ symmetry-breaking
scales of about 150 MeV are near the $\bar{q}q$ \ve, \sr\ mean
mass-splitting scale of $\delta m=141$~MeV, $167$~MeV. However, the CL
$SU(3)$-breaking scale of 18\% for scalars
is about 50\% greater than the 12--13\% scales of \ve, \oca, \de\ states. This
suggests that \ve, \oca, \de\ $\bar{q}q$ or $qqq$ states are all loosely
bound, in contrast with the elementary-particle $\bar{q}q$ \sr\ and, of course, the \ps\ states
(see above). In fact, the latter Nambu--Goldstone \ps\ states are massless in
the CL $p^2=m^2_\pi=0$, $p^2=m^2_K=0$, as the tightly bound
measured \cite{PDG02} $\pi^+$ and $K^+$ charge radii indicate \cite{SKRB03}.

\section{Constituent quarks and the quark-level \lsm}
Formulating the \ps\ and \sr\ $\bar{q}q$ states as elementary chiral partners
\cite{BKRS02}, the Lagrangian density of the $SU(2)$ quark-level linear
$\sigma$ model (\lsm) has, after the spontaneous-symmetry-breaking shift, the
interacting part \cite{GML60} (for $f_\pi=(92.42\pm 0.27)$~MeV~$\approx93$~MeV)
\begin{equation}
{\cal L}^{\mbox{\scriptsize int}}_{\mbox{\scriptsize\lsm}} = g\,\bar{\psi}
(\sigma+i\gamma_5\vec{\tau}\cdot\vec{\pi})\psi\,+\,g'\,\sigma\,(\sigma^2+\pi^2)
\,-\,\frac{\lambda}{4}\,(\sigma^2+\pi^2)^2 \, - \, f_\pi \, g\,\bar{\psi}\psi \; ,
\label{lsm}
\end{equation}
with tree-order CL couplings related as ($f^{\,CL}_\pi\rightarrow 90$ MeV)
\be
g \; = \; \frac{m_q}{f_\pi} \;\;\;\; , \;\;\;\; g' \; = \;
\frac{m^2_\sigma}{2f_\pi} \; = \; \lambda\,f_\pi \; .
\label{ggprime}
\ee
The $SU(2)$ and $SU(3)$ chiral Goldberger--Treiman relations (GTRs) are
\begin{equation}
f_{\pi}\,g \; = \; \hat{m} \;= \; \ha\,(m_u+m_d) \;\;\;\; , \;\;\;\; f_K\,g \;
= \; \ha\,(m_s+\hat{m}) \; .
\label{gtrs}
\end{equation}
Since $f_K/f_\pi\approx1.22$ \cite{PDG02}, the constituent-quark-mass ratio
from \eqr{gtrs} becomes
\be
1.22 \; \approx \; \frac{f_K}{f_\pi} \; = \; \ha\,(1+\frac{m_s}{\hat{m}})
\;\;\;\; \Rightarrow \;\;\;\; \frac{m_s}{\hat{m}} \; \approx \; 1.44 \; ,
\label{rmsmh}
\ee
which is independent of the value of $g$. In loop order, Eqs.~(\ref{ggprime})
are recovered, along with \cite{DS95,SKRB03}
\be
m_\sigma \; = \; 2m_q \;\;\;\; , \;\;\;\; g \; = \; \frac{2\pi}{\sqrt{N_c}}
\;\;\;\; , \;\;\;\; \mbox{for} \;\; N_c \; = \; 3 \; .
\label{loop}
\ee
Here, the first equation is the NJL relation \cite{NJL61}, now true for the
\lsm\ as well. The second equation in Eqs.~(\ref{loop}) was first found via
the $Z=0$ compositeness relation \cite{SWS62_98}, separating the elementary
$\pi$ and $\sigma$ particles from the bound states $\rho$, $\omega$, and $a_1$.

We first estimate the nonstrange and strange constituent
quark masses from the GTRs (\ref{gtrs}), together with the \lsm\ loop-order
result~(\ref{loop}):\footnote{The resulting quark masses are well in agreement with the values obtained on the basis of the magnetic moments of the respective baryons (see e.g.\ Ref.\ \cite{DeRujula:ge}). The proton magnetic moment $\mu_p \simeq 2.7928\;$ e.g.\ yields $\hat{m} = m_p/\mu_p = 336$~MeV.}
\be
\begin{array}{ccccccc}
\hat{m} & \approx & g\,f_\pi & \approx & \displaystyle\frac{2\pi}{\sqrt{3}}\:
(93\:\mbox{MeV}) & \approx & 337\;\mbox{MeV} \quad \stackrel{CL}{\rightarrow} \quad 325\;\mbox{MeV} \; , \\[2mm]
m_s & = & \left(\displaystyle\frac{m_s}{\hat{m}}\right)\hat{m} & \approx
& 1.44\,\hat{m} & \approx & 485\;\mbox{MeV} \quad \stackrel{CL}{\rightarrow} \quad 470\;\mbox{MeV} \; .
\end{array} \label{estim1}
\ee
These quark-mass scales in turn confirm the mass-splitting scales found in
Secs.~2, 3:
\begin{eqnarray}
\delta m_{\cal V}\,\approx\,\delta m_{\cal S} \,\approx\,\delta m_{\!\cal O}
\,\approx\,\delta m_{\!\cal D} & \approx & (485-337)\;\mbox{MeV}\,=\,148\;\mbox{MeV} \nonumber \\
 & \stackrel{CL}{\rightarrow} & (470-325)\;\mbox{MeV}\,=\,145\;\mbox{MeV} \; ,
\label{dmlsm}
\end{eqnarray}
near 141, 167, 150, 150 MeV, respectively. Also the $SU(3)$ non-vanishing
masses are predicted as
\be
\begin{array}{ccccccrcr}
 & & m^0_{\cal V} &\approx& \displaystyle\sqrt{\frac{3}{2}}\;
(m_s+ \hat{m})&\approx& 1007\;\mbox{MeV} \; , & & \\[2mm]
m^0_{\cal O}&=&m^0_{\!\cal D}&\approx&m_s+2\hat{m}&\approx&1160\;\mbox{MeV}\; ,& &
\label{mvaod}
\end{array}
\ee
near the 1048, 1151, and 1232 MeV $m^0$ masses in Secs.~2, 3.

To verify that the pion and kaon are tightly bound $\bar{q}q$ mesons, we
compute the $\pi^+$ and $K^+$ charge radii as \cite{SKRB03}
$r_\pi=1/\hat{m}_{\mbox{\scriptsize CL}}=0.61$ fm and
$r_K=2/(m_s+\hat{m})_{\mbox{\scriptsize CL}}=0.50$ fm, near data \cite{PDG02}
$0.672\pm0.008$ fm and $0.560\pm0.031$ fm, respectively. Likewise, to verify
that the proton is a $qqq$ touching pyramid, we estimate the proton charge
radius as $R_p=(1+\sin\!30^{\circ})\,r_\pi\approx0.9$ fm, near data
\cite{PDG02} $0.870\pm0.008$ fm.

\section{\sr\ scalars and accidental degeneracies}

We begin with the non-CL NJL-\lsm\ scalar masses $m_{\sigma_n}=2\,\hat{m}=674$~MeV, $m_{\kappa}=2\sqrt{\hat{m}\,m_s}=809$ MeV, and $m_{\sigma_s}=2 \,m_s=970$~MeV.

An almost degenerate case in the nonrelativistic quark model (NRQM) is
\cite{MSW88}, in the context of QCD,\footnote{Note that we follow
Ref.~\cite{MSW88}, and use $\alpha_s(m^2_\sigma)\simeq \pi/4$ (see also
Ref.~\cite{Mattingly:ud}), $\vec{L}\cdot \vec{S}=-2$,
$m_{\mbox{\scriptsize dyn}}=315$~MeV, while $\left<r^{-3}\right>=4\,
\beta^3/(3\,\sqrt{\pi}\,)$ is obtained employing harmonic-oscillator wave
functions with $\beta\simeq180$~MeV.}
\be
m_{\!\cal S} \; \approx \; m_{\cal V} \, + \,
\frac{2\alpha_s}{m^2_{\mbox{\scriptsize dyn}}}\,
\left(\frac{\vec{L}\cdot\vec{S}}{r^3}\right) \;
= \; 780\;\mbox{MeV}\,-\,140\;\mbox{MeV} \; = \; 640\;\mbox{MeV} \; ,
\label{nrqm}
\ee
where the ground-state vector mesons have $L=0$ and so no spin-orbit
contribution to the mass.
This corresponds to $m_{\sigma(650)}\approx m_{\omega(782)}-140$ MeV $=642$
MeV. Equivalently, invoking the $I\!=\!1/2$ CGC of $1/2$, one predicts via the
NRQM $\,m_{\kappa(800)}\approx m_{K^*(892)}-70$ MeV $=822$~MeV. Or invoking
instead the $\bar{s}s$ CGC of $1/3$, one gets $m_{\sigma_s(970)}\approx
m_{\phi(1020)}-47$~MeV $= 973$ MeV. In a similar way we obtain also $m_{a_0(985)} = m_{\rho(770)}+(3/2)\, 140\;\mbox{MeV} = 980$~MeV.

As an alternative way to examine the latter, in the case of the
elementary-particle \ps\ and \sr\ states, one should invoke
the infinite-momentum-frame (IMF, see Appendix) scalar-pseudoscalar $SU(3)$
equal-splitting laws (ESLs), reading \cite{S82_92}
\be
m^2_\sigma\,-\,m^2_\pi \; \approx \; m^2_\kappa\,-\,m^2_K \; \approx \;
m^2_{a_0}\,-\,m^2_{\eta_{\mbox{\scriptsize avg}}} \; \approx \; 0.40 \;
\mbox{GeV}^2 \; ,
\label{esls}
\ee
where $m_{\eta_{\mbox{\scriptsize avg}}}$ is the average $\eta$,
$\eta^\prime$ mass 753 MeV. These ESLs hold for the non-CL NJL-\lsm\ scalar mass values. Using
the ESLs~(\ref{esls}) to predict the $a_0$ mass, one finds
\be
m_{a_0}\;=\;\sqrt{0.40\;\mbox{GeV}^2\,+\,m^2_{\eta_{\mbox{\scriptsize avg}}}}
\; \approx \; 983.4\;\mbox{MeV} \; ,
\label{anot}
\ee
very close to the PDG value $984.7\pm1.2$ MeV. Thus, the nearness of the 
$a_0$(985) and $f_0$(980) masses, the latter scalar being mostly $\bar{s}s$
and so near the vector $\bar{s}s$ $\phi$(1020) (see above), is indeed an
accidental degeneracy. Note that a similar (approximate) degeneracy is found
in the dynamical unitarized quark-meson model of \refc{BRMDRR86}, where the
same $\bar{q}q$ assignments are employed as here.

This ground-state scalar $0^+$ nonet [$\sigma(650)$, $\kappa(800)$, $f_0(980)$,
$a_0(985)$] is about 500--700 MeV below the $0^+$ nonet
\cite{PDG02,Kleefeld:2001ds} [$f_0(1370)$, $K_0^*(1430)$, $f_0(1500)$,
$a_0(1450)$], just as the ground-state $1^-$ vector nonet [$\rho(770)$,
$\omega(782)$, $K^*(892)$, $\phi(1020)$] is about 600--800 MeV below the $1^-$
nonet \cite{PDG02} [$\rho(1450)$, $\omega(1420)$, $K^*(1680)$, $\phi(1680)$].

\section{Strong-interaction scalar-meson decay rates} 

Given the above scalar-meson nonet $\sigma(650)$, $\kappa(800)$, $f_0(980)$,
$a_0(985)$, compatible with present data and also with the SU(3) mass
splittings in Secs.~2, 3, 5 and the quark-level L$\sigma$M in Sec.\ 4, we now
predict L$\sigma$M decay rates based on the SU(3) Lagrangian density
${\cal L}^{\mbox{\scriptsize int}}_{\mbox{\scriptsize\lsm}} =
g_{\sigma \pi\pi} \; d_{ijk} \; S_i \;P_j \; P_k$, with L$\sigma$M coupling
$g_{\sigma \pi\pi}=(m^2_\sigma-m^2_\pi)/(2\,f_\pi)\approx 2.18$~GeV, where
$f_\pi= (92.42\pm 0.27)$~MeV and $m_\sigma\approx 650$~MeV (the latter stems
from the CL $m_q\approx 325$~MeV \cite{DS95}).
Thus, the $\sigma\rightarrow 2\,\pi$ decay rate, for
$p_{cm}=294$~MeV and $\phi_s=\pm 18^\circ$\footnote{For convenience, we use
here the same value of the mixing angle $\phi_s$ as in
Ref.~\cite{Kleefeld:2001ds}, i.e., $\phi_s=\pm 18^\circ$.}, becomes
\begin{equation} \Gamma_{\sigma\pi\pi} = \frac{p_{cm}}{8\pi\, m^2_\sigma} \,
\left( \frac{3}{2} \right) \, \left[ \,\,2\, g_{\sigma\pi\pi} \,
\cos\phi_s\,\right]^2 \approx 714\; \mbox{MeV} \; .
\end{equation}
Here the factor of 2 is due to Bose statistics (see e.g.\
Ref.~\cite{Paver:ys}), and this broad width $\Gamma_\sigma \simeq m_\sigma$ is
expected from data \cite{DM2etxy} and from phenomenology \cite{weinetxy}.

Next, the $a_0(985)\rightarrow \eta\pi$ width for $p_{cm} = 321$~MeV is
\begin{equation} \Gamma_{a_0\eta\pi} = \frac{p_{cm}}{8\pi\, m^2_{a_0}} \, \left[ \,\,2\, g_{\sigma\pi\pi} \, \cos\phi_{ps}\,\right]^2 \approx 138 \; \mbox{MeV} \; ,
\end{equation}
where $\phi_{ps} \approx 42^\circ$ is in the quark nonstrange($\bar{n}n$)-strange($\bar{s}s$) basis \cite{KLOEetxy}. This predicted L$\sigma$M width is not incompatible with the high-statistics decay rate \cite{Armstrong:1991rg} $\Gamma_{a_0\eta\pi} = (95 \pm 14)~$MeV.

Furthermore, the $\kappa\rightarrow K\pi$ decay rate, for $p_{cm} = 218$~MeV
and $m_\kappa=800$ MeV, is
\begin{equation} \Gamma_{\kappa K\pi} = \frac{p_{cm}}{8\pi\, m^2_\kappa} \, \left( \frac{3}{4} \right) \, \left[ \,\,2\, g_{\sigma\pi\pi} \,\right]^2 \approx 193 \; \mbox{MeV} \; ,
\end{equation}
which is of the same order as the E791 data \cite{A02}
\begin{equation} \Gamma^{\; \mbox{\scriptsize E791}}_{\kappa K\pi} = (410 \pm 43 \pm 87) \; \mbox{MeV} \; , \quad m_\kappa = (797\pm 19 \pm 43) \; \mbox{MeV}\; ,
\end{equation}
and especially the very recent data of the BES collaboration \cite{Bai:2003fv}
\begin{equation} \Gamma^{\; \mbox{\scriptsize BES}}_{\kappa K\pi} = (220\,{}^{+225}_{-169} \, \pm 97) \; \mbox{MeV} \; , \quad m_\kappa = (771\,{}^{+164}_{-221} \, \pm 55) \; \mbox{MeV}\; .
\end{equation}
Lastly, we estimate (see e.g.\ Ref.\ \cite{Kleefeld:2001ds}) the
$f_0(980)\rightarrow \pi\pi$ rate, assuming again that the $f_0(980)$ is mostly
$\bar{s}s$, with mixing angle $\pm 18^\circ$ in the quark basis
\cite{Scadronetxy}, for $p_{cm}=470$~MeV:
\begin{equation} \Gamma_{f_0\,2\pi} = \frac{p_{cm}}{8\pi\, m^2_{f_0}} \, \left( \frac{3}{2} \right) \, \left[ \,\,2\, g_{\sigma\pi\pi} \,\sin \phi_s \, \right]^2 \approx 53 \; \mbox{MeV} \; ,
\end{equation}
not too distant from the recent E791 measurement \cite{A02}
\begin{equation} \Gamma^{\; \mbox{\scriptsize E791}}_{f_0\,2\pi} = (44 \pm 2 \pm 2) \; \mbox{MeV} \; , \quad m_{f_0} = (977\pm 3 \pm 2) \; \mbox{MeV}\; .
\end{equation}

\section{Electromagnetic meson decay rates involving \bma{\bar{q}q} scalars}
Next we study the five electromagnetic meson decays
$\sigma\rightarrow 2\gamma$, $a_0\rightarrow 2\gamma$,
$f_0\rightarrow 2\gamma$, $\phi\rightarrow f_0\gamma$, and
$\phi\rightarrow a_0\gamma$. Again assuming $m_\sigma \approx 650$~MeV (because
$\hat{m}\approx 325$~MeV $\simeq m_N/3$ in the CL, so that the NJL-L$\sigma$M
scalar mass is $m_\sigma = 2\,\hat{m}\approx 650$~MeV), the quark-loop
amplitude magnitude is, for $f_\pi = (92.42\pm 0.27)$~MeV [20] (see e.g.\
Eq.~(11a) in Ref.~\cite{Karlsen:cm}, and the considerations in
Ref.~\cite{Kleefeld:2001ds})
\begin{equation} |M(\sigma\rightarrow 2\gamma)| \approx \frac{5}{3} \, \frac{\alpha}{\pi\, f_\pi} + \frac{1}{3} \, \frac{\alpha}{\pi\, f_\pi} \approx 5.0 \times 10^{-2}\; \mbox{GeV}^{-1}\, . \label{msgg1}
\end{equation}
Here, the first term is due to the nonstrange quark triangle, while the second
term stems from the charged-kaon and -pion triangle graphs. This result
(\ref{msgg1}) is compatible with the data estimate \cite{Boglione:1998rw}
\begin{equation} \Gamma_{\sigma \, 2\gamma} = \frac{m^3_\sigma}{64\,\pi} \, |M(\sigma\rightarrow 2\gamma)|^2 = (3.8 \pm 1.5)\; \mbox{keV}\, , \end{equation}
or (for $m_{\sigma} \simeq 650$~MeV)
\begin{equation} |M(\sigma\rightarrow 2\gamma)| \simeq (5.3 \pm 1.0)\times 10^{-2}\; \mbox{GeV}^{-1} \, . \end{equation}

Now we examine $a_0(985)\rightarrow 2\gamma$. A nonstrange-quark triangle loop
predicts the gauge-invariant induced amplitude magnitude \cite{Deakin:bm} (for
$m_{a_0}\simeq (984.7 \pm 1.2)$~MeV)
\begin{eqnarray} |M(a_0\rightarrow 2\gamma)|_{\mbox{\scriptsize quark-loop}} & = & \left| 2\, \xi [2+ (1-4\, \xi)\, I(\xi)] \; \frac{\alpha}{\pi f_\pi} \right| \nonumber \\[2mm]
 & = & |\,2.03\pm 0.07 + i \,\,(1.89\pm 0.03)\,| \times 10^{-2}\; \mbox{GeV}^{-1} \nonumber \\[2mm]
 & = & (2.78\pm 0.06) \times 10^{-2}\; \mbox{GeV}^{-1} \, , \label{atgam1}
\end{eqnarray}
for $\xi = \hat{m}^2/m^2_{a_0}\approx 0.109\pm 0.004<1/4$ in the CL, with (see
e.g.\ Ref.~\cite{Kleefeld:2001ds})
\begin{eqnarray} I(\xi) & = & \int_0^1dy \int_0^1 dx \; \frac{y}{\xi - xy (1-y)} \nonumber \\
 & \stackrel{\xi<1/4}{=}& \frac{\pi^2}{2} - 2\, \ln^2 \left[ \frac{1}{\sqrt{4\,\xi}} + \sqrt{\frac{1}{4\,\xi} - 1}\right] + 2\pi i\, \ln \left[ \frac{1}{\sqrt{4\,\xi}} + \sqrt{\frac{1}{4\,\xi} - 1}\right] \nonumber \\[2mm]
 & = &  3.03\pm 0.08 + i \, \,(6.13\pm 0.13)\; .
\end{eqnarray}
However, adding to Eq.~(\ref{atgam1}) the charged-kaon-loop amplitude
\cite{Deakin:bm} $0.97 \times 10^{-2}\; \mbox{GeV}^{-1}$ (as required by the
L$\sigma$M), which has the opposite sign as compared to the fermionic
quark-loop amplitude, in turn predicts \cite{footnote}
\begin{eqnarray} |M(a_0\rightarrow 2\gamma)| & \approx & \left| M(a_0\rightarrow 2\gamma)_{\mbox{\scriptsize quark-loop}} + M(a_0\rightarrow 2\gamma)_{\mbox{\scriptsize kaon-loop}}\right| \nonumber \\[2mm]
 & = & |1.07\pm 0.44 + i\,\,(1.89\pm 0.03) |\times 10^{-2}\; \mbox{GeV}^{-1} \nonumber \\[2mm]
 & = & (2.17\pm 0.22) \times 10^{-2}\; \mbox{GeV}^{-1} \, .
\end{eqnarray}
The latter result is too large as compared to data, assuming $a_0\rightarrow
\eta\pi$ is dominant \cite{PDG02}:
\begin{equation} \Gamma_{a_0 \, 2\gamma} = \frac{m^3_{a_0}}{64\,\pi} \, |M(a_0\rightarrow 2\gamma)|^2 = (0.24 \pm 0.08)\; \mbox{keV}\, , \end{equation}
or
\begin{equation} |M(a_0\rightarrow 2\gamma)| = (0.7 \pm 0.2)\times 10^{-2}\; \mbox{GeV}^{-1} \, . \end{equation}
However, upon disregarding the imaginary part of the quark-loop amplitude,
which is reasonable in view of quark confinement, we come much closer to the
data, as
\begin{equation} \mbox{Re}\Big[ \, M(a_0\rightarrow 2\gamma)_{\mbox{\scriptsize quark-loop}} + M(a_0\rightarrow 2\gamma)_{\mbox{\scriptsize kaon-loop}}\,\Big] = (1.07\pm 0.44) \times 10^{-2}\; \mbox{GeV}^{-1} \; .
\end{equation}

Next we study $f_0\rightarrow 2\, \gamma$. Assuming for the moment that
$f_0(980)$ is purely $\bar{s}s$, the strange-quark loop gives, for $N_c=3$
\cite{Delbourgoetxy} (see also Ref.~\cite{Kleefeld:2001ds})
\begin{equation} |M(f_0\rightarrow 2\gamma)|_{\mbox{\scriptsize quark-loop}} =  \frac{\alpha \;N_c \; g_{f_0\,SS}}{9\,\pi\; m_s} \simeq 8.19 \times 10^{-3}\; \mbox{GeV}^{-1} \, , \label{fftgg1}
\end{equation}
taking the L$\sigma$M value $m_s=485$~MeV from Eq.\ (\ref{estim1}),
with the L$\sigma$M coupling $g_{f_0\,SS} = 2\pi \sqrt{2/3} \approx 5.13$. 
In fact, Eq.~(\ref{fftgg1}) is surprisingly near the observed amplitude
\cite{PDG02}
\begin{equation} \Gamma_{f_0 \, 2\gamma} = \frac{m^3_{f_0}}{64\,\pi} \, |M(f_0\rightarrow 2\gamma)|^2 = (0.39 \pm 0.12)\; \mbox{keV}\, , \end{equation}
or (with $m_{f_0}\simeq (980\pm 10)$~MeV)
\begin{equation} |M(f_0\rightarrow 2\gamma)| = (9.1 \pm 1.5)\times 10^{-3}\; \mbox{GeV}^{-1} \, . \end{equation}
Nevertheless, a more detailed analysis based on kaon and pion loops,
and allowing a small $\bar{n}n$ admixture in the $f_0(980)$, essentially
confirms this nice result \cite{Kleefeld:2001ds}).

Let us now analyse the decay $\phi(1020)\rightarrow f_0(980) \gamma$. Since the
$\phi(1020)$ is known to be dominantly $\bar{s}s$, just as we assume the
$f_0(980)$ to be, the $s$-quark loop gives (with $g_\phi=13.43$ from
 $\Gamma_{\phi ee}$ and $e=\sqrt{4\pi\,\alpha}=0.30282\ldots$)
\begin{equation} |M(\phi\rightarrow f_0\gamma)|_{\mbox{\scriptsize quark-loop}} =  \frac{2\, g_\phi\, e \; g_{f_0\,SS}}{4\,\pi^2\; m_s} \; \cos\phi_s \simeq 2.07 \; \mbox{GeV}^{-1} \, . \label{phifgq1}
\end{equation}
However, the charged-kaon loop is known to give the rate \cite{LucioMartinez:uw}
\begin{equation} \left.\Gamma_{\phi f_0 \,\gamma}\,  \right|_{\mbox{\scriptsize kaon-loop}} = 8.59\times 10^{-4}\; \mbox{MeV}\, , \end{equation}
or
\begin{equation} |M(\phi\rightarrow f_0 \, \gamma)|_{\mbox{\scriptsize kaon-loop}} = 0.75\; \mbox{GeV}^{-1} \, . \label{phifgam1}\end{equation}
Subtracting this kaon-loop amplitude (\ref{phifgam1}) from the quark-loop amplitude (\ref{phifgq1}) predicts in turn
\begin{equation} |M(\phi\rightarrow f_0 \, \gamma)| \approx 2.07 \; \mbox{GeV}^{-1} - 0.75 \; \mbox{GeV}^{-1} = 1.32 \; \mbox{GeV}^{-1} \; , \end{equation}
near the recent KLOE data \cite{Aloisio:2002bt}, for $p_{cm}\simeq(38.69\pm
9.62)$~MeV,
\begin{equation} \left.\Gamma_{\phi f_0 \,\gamma} \, \right|_{\mbox{\scriptsize KLOE}} = \frac{p^3_{cm}}{12 \pi} \; |M(\phi\rightarrow f_0 \, \gamma)|^2 \approx  (19\pm 1) \times 10^{-4}\; \mbox{MeV}\, , \end{equation}
or
\begin{equation} |M(\phi\rightarrow f_0 \, \gamma)| \approx (1.11 \pm 0.42)\; \mbox{GeV}^{-1} \, , \end{equation}
as the branching rate for $\phi\rightarrow f_0 \,\gamma$ is
$(4.47\pm 0.21)\times 10^{-4}$.

Lastly we note that the KLOE observed branching ratio (BR) is
\cite{Aloisioetxy}
\begin{equation} \mbox{BR}(\phi \rightarrow f_0 \gamma / a_0 \gamma) = 6.1 \pm 0.6 \; . \label{branrat1}
\end{equation}
Because we know that $\phi$ is dominantly $\bar{s}s$, this BR Eq.\ (\ref{branrat1}) being much greater than unity strongly suggests that $a_0(985)$ is mostly $\bar{n}n$ and $f_0(980)$ is mostly $\bar{s}s$. The latter assumption we have
continually made throughout this paper, while it had been a conclusion of
Ref.~\cite{Kleefeld:2001ds} (see also Ref.\ \cite{Anisovich:2003up}).

\section{\bma{W}-emission weak decay rates}
In this section we study the five weak decays $K^+\rightarrow\pi^0\,\pi^+$,
$D^+\rightarrow \overline{K}^0\,\pi^+$, $D^+\rightarrow\sigma\,\pi^+$,
$D^+\rightarrow\pi^0\,\pi^+$, and $D_s \rightarrow f_0(980) \,\pi^+$, via 
tree-level $W$-emission graphs. Recalling from Refs.~\cite{Kleefeld:2001ds} and
\cite{Anisovich:2003up}, the amplitudes due to $W$ emission are\footnote
{We use here \mbox{$G_F = 1.16639(1)\times 10^{-5}\;\mbox{GeV}^{-2}$},
$|V_{ud}|=0.9735\pm 0.0008$, $|V_{us}|=0.2196\pm 0.0023$, $|V_{cd}|=0.224\pm
0.016$, $|V_{cs}|=1.04\pm 0.16$, $m_{D^+}=(1869.4\pm 0.5)$~MeV, and
$m_{D_s^+}=(1969.0\pm 1.4)$~MeV.}, for $f_\pi = (92.42\pm 0.27)$~MeV,
\begin{eqnarray} |M(K^+\rightarrow \pi^0\,\pi^+)| & = & 
\frac{G_F \, | V_{ud} | \, | V_{us} |}{2\sqrt{2}} \; f_\pi \;
(m^2_{K^+}-m^2_{\pi^0}) \nonumber \\[2mm]
 & = & (1.837 \pm 0.020) \cdot 10^{-8}\;
\mbox{GeV} \; , 
\end{eqnarray}
near data \cite{PDG02} $(1.832 \pm 0.007) \cdot 10^{-8}\; \mbox{GeV}$, 
\begin{eqnarray}
|M(D^+\rightarrow \bar{K}^0\,\pi^+)| & = & 
 \frac{G_F \, | V_{ud} | \, | V_{cs} |}{2} \; f_\pi \;
(m^2_{D^+}-m^2_{\bar{K}^0}) \nonumber \\[2mm] 
 & = & (177 \pm 27) \cdot 10^{-8}\; \mbox{GeV} \; , 
\end{eqnarray}
near data \cite{PDG02} $(136 \pm 6) \cdot 10^{-8}\; \mbox{GeV}$, and
\begin{eqnarray}
|M(D_s^+\rightarrow f_0\,\pi^+)| & = & 
\frac{G_F \, | V_{ud} | \, | V_{cs} |}{2} \; f_\pi \;
(m^2_{D_s^+}-m^2_{f_0}) \nonumber \\[2mm]
 & = & (159 \pm 25) \cdot 10^{-8}\; \mbox{GeV} \; ,
\end{eqnarray}
near data \cite{PDG02} $(178 \pm 40) \cdot 10^{-8}\; \mbox{GeV}$. In the latter case we have assumed that $f_0(980)$ is all $\bar{s}s$.

Now we also consider $D\rightarrow \pi^0\,\pi^+$ and
$D\rightarrow \sigma\,\pi^+$ (with $m_\sigma=650$~MeV), again in this
$W$-emission scheme, predicting 
\begin{eqnarray}
|M(D^+\rightarrow \pi^0\,\pi^+)| & = & 
\frac{G_F \, | V_{ud} | \, | V_{cd} |}{2\sqrt{2}} \; f_\pi \;
(m^2_{D^+}-m^2_{\pi^0}) \nonumber \\[2mm]
 & = & (28.9 \pm 2.1) \cdot 10^{-8}\; \mbox{GeV} \; ,
\end{eqnarray}
near data \cite{PDG02} $(38.6 \pm 5.4) \cdot 10^{-8}\; \mbox{GeV}$ (also see
Ref.~\cite{mikeeg}, with $p_{cm}=925$~MeV), and
\begin{eqnarray}
|M(D^+\rightarrow \sigma\,\pi^+)| & = & 
\frac{G_F \, | V_{ud} | \, | V_{cd} |}{2\sqrt{2}} \; f_\pi \;
(m^2_{D^+}-m^2_{\sigma}) \nonumber \\[2mm] 
 & \simeq &  25.5 \cdot 10^{-8}\; \mbox{GeV} \; ,
\end{eqnarray}
near\footnote{At this point we should keep in mind that the uncertainty in $m_\sigma$ is of the order of $m_\sigma$!} recent data \cite{PDG02} $(37.6 \pm 4.5) \cdot 10^{-8}\; \mbox{GeV}$. This latter amplitude follows from the decay rate (with $p_{cm}=815$~MeV, $\tau=1051\times 10^{-15}$~s)
\begin{eqnarray} \Gamma_{D^+\sigma\pi^+} & = & \frac{p_{cm}}{8\pi\,m^2_{D^+}} \; |M(D^+\rightarrow \sigma\,\pi^+)|^2 \nonumber \\
 & = & \frac{h}{2\pi\,\tau} \; (2.1\pm 0.5) \times 10^{-3} = (1.32 \pm 0.31) \times 10^{-15} \; \mbox{GeV} \; .
\end{eqnarray}
Not only are the above $D^+\rightarrow \pi^0\,\pi^+$ and $D^+\rightarrow
\sigma\,\pi^+$ $W$-emission amplitudes near data, they are even of about the
same magnitude. This is another example of the $\sigma$ and $\pi$ being chiral
partners \cite{BKRS02}.

\section{Summary and conclusions}
Throughout this paper we have dealt with all ground-state mesons as $\bar{q}q$
nonets in the context of the L$\sigma$M. In Sec.~2 we studied SU(3) mass
splittings for \ve\ and \sr\ $\bar{q}q$ mesons, with \ve\ loosely bound states,
and \ps, \sr\ tighter $\bar{q}q$ elementary particles. In Sec.~3 we reviewed
$qqq$ octet and decuplet baryons. In Sec. 4 we briefly summarized the
quark-level L$\sigma$M theory, while in Sec.\ 5 we explained the accidental
degeneracy of the $a_0(985)$ and $f_0(980)$ scalars. In Sec.~6 we computed
a few strong scalar-meson decay widths, while in Sec.\ 7 we performed a similar
analysis for some electromagnetic decays involving scalar mesons. Finally, in
Sec.\ 8 we employed $W$-emission graphs to describe several hadronic
weak-decay processes.

The usual field-theory picture is that meson masses should appear quadratically
and baryon masses linearly in Lagrangian models based on the Klein--Gordon
and Dirac equations. However, in Secs.~2 and 3 we studied both mesons and
baryons in a {\it linear}-mass $SU(3)$-symmetry Taylor-series sense. Instead,
in Sec.~5 we studied symmetry breaking in the IMF, with
$E=[p^2+m^2]^{1/2}\approx p\,[1+m^2/2p^2+\ldots]$. Here, between brackets, the
$1$ indicates the symmetry limit, and the \em quadratic \em \/mass term means
that both meson and baryon masses are \em squared \em \/in the mass-breaking
IMF for $\Delta S\!=\!1$ ESLs. While the former mass-splitting approach (with
linear masses) fits all \ve, \sr, \oca, and \de\ ground-state
$SU(3)$-flavor multiplets, so does the latter (with quadratic masses) for the
IMF-ESLs. Nevertheless, Nambu--Goldstone pseudoscalars \ps\ \em always \em
\/involve \em quadratic \em \/masses. Both approaches suggest that all 
ground-state mesons (\ps, \sr, \ve) are $\bar{q}q$ states, while baryons
(\oca, \de) are $qqq$ states. This picture is manifest in the quark-level \lsm\
of Sec.~4. The accidental scalar degeneracy between the $\bar{s}s$
$f_0$(980) and the $\bar{n}n$ $a_0$(985) was explained in Sec.~5, via the IMF
quadratic-mass ESLs --- also compatible with mesons being $\bar{q}q$ and
baryons $qqq$ states.

Concerning the mass splittings in general, we observed the remarkable feature
that the real parts of masses of resonances in mesonic and baryonic
ground-state multiplets nicely follow an SU(3) splitting pattern, despite
the enormous disparities in decay widths and thus in the imaginary parts. This
may be understood in the unitarized picture of Ref.~\cite{BRMDRR86}, in which
both real and virtual decay channels contribute to the physical masses of e.g.\
the scalar mesons as dressed $\bar{q}q$ states. We also verified in
Secs.~6, 7, and 8 that mesonic decay rates can be simply explained on the basis
of the flavor and chiral symmetry underlying the quark-level L$\sigma$M. This
is another indication that the lowest lying mesons are all $\bar{q}q$, while
the considered baryons are $qqq$. 

So far we have taken the mass and coupling parameters of the quark-level \lsm\
--- in particular $m_\sigma$ --- to be real numbers (``narrow-width
approximation''). A recently developed formalism \cite{kleefeldxy} may allow us
to go beyond this approximation in the near future. \\[5mm]

\noindent{\large\bf Acknowledgments} \\[1mm]
One of the authors (M.D.S.) wishes to thank V.~Elias and N.~Paver for useful
discussions.
This work was partly supported by the
{\it Funda\c{c}\~{a}o para a Ci\^{e}ncia e a Tecnologia}
of the {\it Minist\'{e}rio da Ci\^{e}ncia e da Tecnologia} \/of Portugal,
under contract no.\ POCTI/\-FNU/\-49555/\-2002, and grant no.\
SFRH/\-BPD/\-9480/\-2002.

\appendix

\section{Kinematic infinite-momentum frame}

The infinite-momentum frame (IMF) has two virtues: \,(i) 
$\,E=[p^2+m^2]^{1/2}\approx p+m^2/2p+\ldots$, for $p\to\infty$, requires \em
squared \em \/masses when the lead term $p$ is eliminated, using $SU(3)$
formulae with coefficients $1\!+\!3=2\!+\!2$, as e.g.\ the Gell-Mann--Okubo
linear mass formula $\Sigma\!+\!3\Lambda=2N\!+\!2\Xi$, valid to 3\%; \,(ii)
when $p\to\infty$, dynamical tadpole graphs are suppressed \cite{FF65}. In
fact, $\Sigma^2\!+\!3\Lambda^2=2N^2\!+\!2\Xi^2$ is also valid empirically to
3\%.  This squared $qqq$ baryon mass formula can be interpreted as a
$\Delta S\!=\!1$ ESL, which holds for both \oca\ and \de\ baryons \cite{S82_92}:
\be
\begin{array}{ccccccccc}
\Sigma\Lambda-N^2 &\!\!\approx\!\!& \Xi^2-\Sigma\Lambda &\!\!\approx\!\! &
\ds\ha\left(\Xi^2-N^2\right) & & &\!\!\approx\; 0.43\;\mbox{GeV}^2 \; , \\[2mm]
{\Sigma^*}^2-\Delta^2 &\!\!\approx\!\!& {\Xi^*}^2-{\Sigma^*}^2 &\!\!\approx\!\!
&\Omega^2-{\Xi^*}^2&\!\!\approx\!\!& \ds\ha\,\left(\Omega^2-{\Sigma^*}^2\right)
&\!\!\approx \;0.43\;\mbox{GeV}^2 \; .
\end{array}
\label{baresls}
\ee

However, the $\bar{q}q$ pseudoscalar and vector $\Delta S\!=\!1$ ESLs have
about one half this scale (also empirically valid to 3\%), viz.\
\be
m^2_K\,-\,m^2_\pi \; \approx \; m^2_{K^*}\,-\,m^2_\rho \; \approx \;
m^2_\phi\,-\,m^2_{K^*}\; \approx \; \ha\,(m^2_\phi\,-\,m^2_\rho) \; \approx \;
0.22\;\mbox{GeV}^2 \; ,
\label{pseudoscalars}
\ee
as roughly do the $\bar{q}q$ scalars found in Sec.~2, i.e.,
\be
m^2_{\kappa(800)}\,-\,m^2_{\sigma_{n}(650)} \; \approx \; m^2_{\sigma_s(940)}
\,-\,m^2_{\kappa(800)} \; \approx \; \mbox{0.22\,\ldots\,0.24 GeV}^2 \; .
\label{scalars}
\ee
This approximate factor of 2 between Eqs.~(\ref{baresls}) and
Eqs.~(\ref{pseudoscalars},\ref{scalars}) is because there are two
$\Delta S\!=\!1$ $qqq$ transitions, whereas there is only one $\Delta S\!=\!1$
transition for $\bar{q}q$ configurations.

So if we take \eqr{scalars} as physically meaningful, we may write
\be
2m^2_{\kappa} \; \approx \; m^2_{\sigma(600)} \,+\,m^2_{f_0(980)} \; \approx \;
m^2_{\sigma_{n}(650)} \,+\,m^2_{\sigma_{s}(940)} \; \approx \;
\mbox{1.31\,\ldots\,1.32 GeV}^2 \; ,
\label{kappa}
\ee
yielding $m_\kappa\approx811$ MeV close to experiment, which again suggests
these scalars are $\bar{q}q$ states.

These IMF quadratic mass schemes, along with the non-CL NJL-\lsm\ $\kappa$ mass
$m_{\kappa(809)}=2\sqrt{\hat{m}\,m_s}=809$ MeV or the averaged\footnote{We average here approximately between the non-CL NJL-\lsm\ mass value and the respective value in the CL.} mass value of 800 MeV, again suggest (as do the
empirical scales of \eqrs{pseudoscalars}{scalars} vs. Eqs.~(\ref{baresls}))
that \em all \em \/ground-state meson nonets are $\bar{q}q$, whereas the baryon
octet and decuplet are $qqq$ states.


\begin{thebibliography}{99}
\bibitem{footnote}
Recall that the charged-kaon-loop amplitude is obtained by
\begin{eqnarray*} M(a_0\rightarrow 2\gamma)_{\mbox{\scriptsize kaon-loop}} & = &
\frac{2\,\alpha}{\pi\,m^2_{a_0}} \, \frac{m^2_{a_0} - m^2_{K^\pm}}{2\, f_K} \,
\left[ - \frac{1}{2} + \xi_K \,\, I(\xi_K) \right]\, .
\end{eqnarray*}
With $\xi_K = m_{K^\pm}/m^2_{a_0}\approx 0.251 \pm 0.008 >1/4$ and $f_K=(113\pm
1.04)$~MeV, we get
\begin{eqnarray*} I(\xi_K) & \stackrel{\xi_K>1/4}{=}& 2\,
\arcsin^2 \left(\sqrt{4\,\xi_K}\right)^{-1} \quad = \quad 4.48 \pm 1.25 \; ,
\end{eqnarray*}
and $M(a_0\rightarrow 2\gamma)_{\mbox{\scriptsize kaon-loop}} =
(0.97\pm 0.43)\times 10^{-2}\; \mbox{GeV}^{-1}$.
\end{thebibliography}
\end{document}